\documentclass[]{tMPH2e}

\usepackage[english]{babel}
\usepackage{lmodern}

\usepackage{multirow}
\usepackage[version=3]{mhchem} 
\usepackage{amsmath}
\usepackage{amssymb}
\usepackage{threeparttable,array,booktabs,calc}

\usepackage{rotating}

\begin{document}

\doi{10.1080/0026897YYxxxxxxxx}
\issn{13623028}
\issnp{00268976}
\jvol{00}
\jnum{00} \jyear{2016} 

\articletype{ARTICLE}

\title{Spin--Orbit Coupling and Rovibrational Structure in the Iododiacetylene Radical Cation by PFI-ZEKE Photoelectron Spectroscopy}

\author{
\vspace{6pt} Katrin Dulitz$^{a}$, Elias Bommeli$^{a}$, Guido Grassi$^{a}$, Daniel Zindel$^{a}$, and Fr\'ed\'eric Merkt$^{a}$$^{\ast}$\thanks{$^\ast$Corresponding author. Email: merkt@phys.chem.ethz.ch
\vspace{6pt}}\\
\vspace{6pt} $^{a}${\em{ETH Z\"urich, Laboratorium f\"ur Physikalische Chemie, Vladimir-Prelog Weg 2, CH-8093 Z\"urich, Switzerland}}\\
}
\maketitle

\begin{abstract}
The photoelectron spectrum of the $\textrm{X}^{+}\,{}^{2}\Pi \leftarrow \textrm{X}\,{}^{1}\Sigma^{+}$ photoionising transition in iododiacetylene, \ce{HC4I}, has been recorded using pulsed-field-ionisation zero-kinetic-energy (PFI-ZEKE) photoelectron spectroscopy with partial resolution of the rotational structure. The first adiabatic ionisation energy of \ce{HC4I} and the spin--orbit splitting of the X$^{+}\,{}^{2}\Pi$ state of \ce{HC4I+} are determined as $E^{\textrm{ad}}_{\textrm{I}}/(hc) = 74470.7(2)$~cm$^{-1}$ and $\Delta\tilde{\nu}_{\textrm{so}} = 1916.7(4)$~cm$^{-1}$, respectively. Several vibrational levels of the X$^{+}\,{}^{2}\Pi$ electronic ground state of the \ce{HC4I+} cation have been observed. The experimental data are discussed in the realm of a simple three-state charge-transfer model without adjustable parameters which allows for a qualitative description of the electronic structure and spin--orbit coupling in \ce{HC4I+} and of the change in bond lengths upon ionisation of HC$_4$I.\bigskip

\end{abstract}

\section{\label{sec:introduction} Introduction}

Haloacetylenes, \mbox{H--C$\equiv$C--X}, and halodiacetylenes, \mbox{H--C$\equiv$C--C$\equiv$C--X} (with X $=$ F, Cl, Br, I), are of interest for the investigation of charge migration processes in molecules, because they contain conjugated $\pi$-type molecular orbitals along the (C$\equiv$C)$_2$ molecular axis and p$_{x,y}({\rm p}_\pi)$-type atomic orbitals at the halogen end group. The presence of a halogen substituent affects the electron-density distribution along the carbon chain through two counteracting effects. On the one hand, the halogen atoms (in the order F~$>$~Cl~$>$~Br~$>$~I) lead to an inductive, electron-withdrawing effect from the carbon chain. On the other hand, the p$_{\pi}$-orbitals of the halogen substituent can contribute to the conjugated $\pi$-electron system of the chain, thus strengthening the resonance interaction within the molecule \cite{Uneyama2007}. Photoelectron spectroscopy is well suited to study the charge distribution in $^2\Pi$ electronic states of positively charged carbon chains with halogen end groups because the observed spin--orbit splittings are directly proportional to the ${\rm p}_\pi$ electron-hole density at the halogen atom \cite{Haink1970, Heilbronner1970, Heilbronner1974, Allan1977, Maier1980, Gruetter2012, Gans2013}. 

He I photoelectron spectra of halogen-substituted acetylenes and diacetylenes \cite{Haink1970, Heilbronner1970, Heilbronner1974, Allan1977, Maier1980} and the electronic spectra of the corresponding cations \cite{Allan1977, Allan1978, Allan1979, Maier1979, Maier1980, Maier1981, Maier1982a, Klapstein1982, Klapstein1983, Leutwyler1983, Leutwyler1984, Klapstein1986} reveal an increase of the spin--orbit splitting of the electronic ground state of the cations in the order F~$<$~Cl~$<$~Br~$<$~I, in accordance with the strength of the spin--orbit coupling observed in the $^2{\rm P}$ ground state of the halogen atoms. 

In their systematic investigations of the He I photoelectron spectra of (di)haloacetylenes and (di)halodiacetylenes, Heilbronner, Maier and their coworkers \cite{Haink1970, Heilbronner1970, Heilbronner1974, Allan1977, Maier1980} have measured the positions and spin--orbit splittings of all low-lying electronic states of the cations. They also derived a semi-quantitative model to describe the observed spectra on the basis of an LCBO (linear combination of bond orbitals) approximation  including the coupled lone-pair p$_\pi$ orbitals of the halogen atoms and the $\pi$(CC) bonding orbitals of the acetylenic groups. The model provided a satisfactory description of the measured electronic intervals and spin--orbit splittings for the halo(di)acetylenes after adjustment of the model parameters \cite{Heilbronner1974}. However, it failed to account for the spin--orbit splittings observed in the spectra of the dihalodiacetylenes in a consistent manner and needed to be extended by inclusion of the $\pi^*$(CC) antibonding orbitals, which made the model formally equivalent to the standard H\"uckel molecular-orbital treatment.
Minimal orbital sets, in spite of their obvious shortcomings, help extracting or predicting the main features of photoelectron spectra and of the corresponding electronic structure and dynamics and so provide a simple link between time- and frequency-resolved measurements of charge delocalisation and migration, including the effects of spin--orbit coupling. 

In a previous study, the rotationally resolved photoelectron spectrum of iodoacetylene, \ce{HC2I}, was obtained using pulsed-field-ionisation zero-kine\-tic-ener\-gy (PFI-ZEKE) photoelectron spectroscopy \cite{Gans2013}. In the X$^{+}\,{}^{2}\Pi$ electronic ground state of the \ce{HC2I+} cation, the spin--orbit coupling is much stronger than the Renner--Teller effect and effectively quenches vibronic interactions. A two-state charge-transfer model with only three parameters -- the known vertical ionisation energy $E_{\textrm{I,HCCH}}$ of acetylene, the known ionisation energy $E_{\textrm{I,X}}$ of X, and an adjustable charge-transfer coupling $V_{\rm X-CC}$ between the p$_\pi$ lone-pair orbitals of the halogen atom and the acetylenic $\pi$ orbitals -- was applied to describe the lowest two electronic states of \ce{HC2X+} and provided a semi-quantitative description of the structure of the photoelectron spectra. Recently, Kraus et al. used high-harmonic spectroscopy to measure the charge-migration dynamics in the time domain following ionisation of \ce{HC2I} with an ultrashort laser pulse \cite{Kraus2015}.

Here, we report on the measurement of the PFI-ZEKE photoelectron spectrum of iododiacetylene, \ce{HC4I}, at a resolution sufficient to partially resolve the rotational structure and to determine precise values of the ionisation energies and spin--orbit splittings. Compared to \ce{HC2I}, \ce{HC4I} possesses an additional C$\equiv$C triple bond and can thus be used to probe the effect of an extended $\pi$-system on the photoionisation process in a halogen-substituted, linear acetylenic molecule. 

For the halodiacetylene cations, the minimal orbital set includes p$_\pi$ lone-pair orbitals of the halogen atoms and the acetylenic $\pi$ orbitals. The corresponding three-state model can be described by the matrix (all elements are chosen to be real)
\begin{equation}
\textbf{H} =
\begin{pmatrix}
E_{\textrm{I,X}}  & hcV_{\textrm{X-CC}}            & 0 \\
hcV_{\textrm{X-CC}}    & E_{\textrm{I,\ce{HC2H}}}  & hcV_{\pi\pi} \\
0                 & hcV_{\pi\pi}   & E_{\textrm{I,\ce{HC2H}}} \\
\end{pmatrix}
.\label{eq:Hamiltonian}
\end{equation}
With this model, the structure of the lowest three $^2\Pi$ states of \ce{HC4X+} can be described by the same three parameters as \ce{HC2X+}, and only one additional parameter, the charge-transfer coupling parameter $V_{\pi\pi}$ between the two acetylenic $\pi$ systems. The interaction term $V_{\pi\pi}$ can be obtained independently from the photoelectron spectrum of diacetylene \cite{Smith1967, Brogli1973}. In the case of \ce{HC4X+}, it is thus possible to predict the main features of the electronic structure using equation~(\ref{eq:Hamiltonian}) without any adjustable parameter. The spin--orbit splittings of the three low-lying $^2\Pi$ electronic states of HC$_4$I$^+$ are obtained by multiplying the spin--orbit coupling constant of the halogen atom with $|c_{\textrm{X},i}|^2$, where $c_{\textrm{X},i}$ is the coefficient describing the contribution of the halogen-atom p$_\pi$ orbital to the eigenfunction of the $i$-th electronic state. Because the Renner--Teller effect and the spin--orbit interaction are in competition \cite{Barckholtz1998,Gans2013}, the magnitude of the spin--orbit splitting enables one to assess whether the Renner--Teller effect is likely or not to be quenched by the spin--orbit interaction. Large values of $|c_{\textrm{X},i}|^2$ also indicate a shortening of the C--I bond, which, in turn, leads to an increase of the rotational constant compared to the neutral ground state \cite{Gans2013}. 

These considerations, though rather simple and qualitative, provide an attractive realm to discuss the results of spectroscopic investigations.
Anticipating the results presented in the next sections, the experimentally observed adiabatic ionisation energy of HC$_4$I is found to be lower than that of HC$_2$I, and the spin--orbit splitting of the X$^{+}\,{}^{2}\Pi$ ground electronic state is found to be smaller than in the ground state of \ce{HC2I+}. However, the spin--orbit splitting remains large enough to suppress observable effects of the Renner--Teller effect. Finally, the rotational constant of the ground vibronic state of \ce{HC4I+} is found to be larger than in the neutral ground state.

This article has the following structure: After a short description of the experimental setup and procedure in Section \ref{expsetup}, Section \ref{results} presents the photoelectron spectra and their analysis. In the last part of the article, the ionisation energies, spin--orbit splittings and rotational constants extracted from the analysis are discussed in the realm of the qualitative three-state model just described.

\section{\label{sec:experimental} Experiment}\label{expsetup}

The measurements were carried out using a narrow-bandwidth (0.008~cm$^{-1}$) vacuum-ultraviolet (VUV) laser system and a PFI-ZEKE photoelectron spectrometer described in earlier work \cite{Hollenstein2000, Hollenstein2001}. Here, we only give a brief description of the experimental setup and emphasize the aspects specific to the present measurements. The chemical synthesis of \ce{HC4I}, adapted from procedures described in Refs.~\cite{KlosterJensen1966,Gans2013}, is described in the appendix.

VUV laser radiation is generated by resonance-enhanced difference-frequency mixing ($\nu_{\textrm{VUV}} = 2\nu_{\textrm{UV}}-\nu_2$) in \ce{Xe} and \ce{Kr} gas, respectively. Laser radiation of frequencies $\nu_{\textrm{UV}}$ and $\nu_2$ is produced separately by two continuous-wave (cw) ring dye lasers and subsequent pulsed-dye amplification (PDA). An injection-seeded Nd:YAG laser (Quanta-Ray Lab-170, 532~nm, 25~Hz repetition rate, pulse energies attenuated to $\approx$~30~mJ) is used to pump the two PDA lines. The ring dye lasers are typically pumped by 6-7~W of 532~nm laser radiation from cw \ce{Nd}:\ce{YVO4} lasers. One of the ring dye lasers (Coherent 699) is held at a fixed frequency corresponding to $\nu_{\textrm{UV}}/3$ (Sulforhodamine 101 dye) in order to generate -- after pulsed dye amplification (dye mixture of DCM and Pyridine 1) and frequency upconversion in two succesive BBO crystals -- the UV laser radiation required to excite a two-photon resonance in \ce{Xe} and \ce{Kr} gas at the wave numbers $2\tilde{\nu}_{\textrm{UV}} = 89860.02$~cm$^{-1}$ and $2\tilde{\nu}_{\textrm{UV}} = 92307.38$~cm$^{-1}$\  \cite{Miyazaki1989, NIST_ASD}, respectively. The fundamental wave number of the ring dye laser is stable to within 0.01~cm$^{-1}$. The second, frequency-tunable ring dye laser (Coherent 899-21 Autoscan II) is operated with the dye Kiton Red to produce laser radiation in the wave-number range $\tilde{\nu}_2 = 15250 - 16100$~cm$^{-1}$. The wave number $\tilde{\nu}_2$ is calibrated by simultaneous detection of the fluorescence of molecular iodine (I$_2$) induced by deflecting $\approx 5$~\% of the laser beam through an \ce{I2} vapour cell. The PDA beam of frequency $\nu_2$ is overlapped with the other laser beam ($\nu_{\textrm{UV}}$) using a dichroic mirror. VUV laser radiation is produced inside a vacuum chamber by resonance-enhanced four-wave mixing just below the orifice of a pulsed valve through which \ce{Xe} or \ce{Kr} gas (2~bar stagnation pressure) is expanded. A diffraction grating is used to separate the VUV radiation ($\nu_{\textrm{VUV}}$) from laser light of other frequencies, e.g. $\nu_{\textrm{UV}}$ and $\nu_{2}$, and to deflect it into a probe-gas chamber, where it is used to excite $\ce{HC4I}$ molecules to the region near the first adiabatic ionisation threshold.

A mixture of \ce{HC4I} in \ce{He} carrier gas is produced by passing \ce{He} gas (1~bar) through a stainless-steel cylinder containing a small quantity of liquid \ce{HC4I}. In this way, gaseous \ce{HC4I} (5~mbar vapour pressure at 293~K\  \cite{KlosterJensen1966}) is entrained into the gas flow and expanded into vacuum through a pulsed solenoid valve (General Valve, Series 9). Over time, decomposition of \ce{HC4I} in the valve body leads to the formation of a soot-like residue which eventually blocks the valve orifice.
A skimmer (Beam Dynamics, 2~mm diameter orifice) separates the expansion region from a zone in which VUV laser excitation, pulsed-field-ionisation and electron detection is carried out.

The PFI-ZEKE photoelectron spectra are recorded by monitoring the electrons produced by delayed pulsed field ionisation of high Rydberg states (principal quantum number $n>200$) \cite{MuellerDethlefs1998}. To achieve a high spectral resolution in the PFI-ZEKE photoelectron spectrum, a multipulse field-ionisation sequence is used as detailed in Ref. \cite{Hollenstein2001}. For the detection of the X$^+$ $^2\Pi$ state of \ce{HC4I+}, a positive discrimination pulse (500~ns duration, 125~mV/cm) is followed by a series of successive negative pulses of increasing amplitude. Time gates are set at the positions corresponding to the times of flight of the electrons produced by these pulses and the integrated electron signals are stored as a function of the VUV wave number. The electric-field-induced shift of the ionisation thresholds is corrected for as described previously \cite{Hollenstein2001}.

\section{Results}
\label{results}
\subsection{Rotational Structure and Spin--Orbit Coupling in the X$^{+}\,{}^{2}\Pi$ State}

In the electronic ground state of both the neutral (X$\,{}^{1}\Sigma^{+}$) and the cation (X$^{+}\ {}^{2}\Pi$), iododiacetylene is assumed to be a linear molecule with C${_{\infty \textrm{v}}}$ symmetry. Owing to the presence of the iodine atom, which exhibits a strong spin--orbit coupling (coupling constant $\tilde{A}_{\textrm{I}} = 2(E_{3/2}-E_{1/2})/(3hc)= -5069$~cm$^{-1}~$ \cite{NIST_ASD}), the molecule is best treated in a Hund's case (a) basis. In the following, we use a double prime ($''$) and a superscript plus sign (${}^{+}$) when we refer to the quantum numbers of \ce{HC4I} and \ce{HC4I+}, respectively. The photoelectron spectrum of the $\textrm{X}^{+}\,{}^{2}\Pi_{{\Omega}^{+}} \leftarrow \textrm{X}\,{}^{1}\Sigma^{+}$ photoionising transition of \ce{HC4I} consists of two distinct spin--orbit components with ${\Omega}^{+} =\left|{\Lambda^{+}}+{\Sigma^{+}}\right| = 1/2$ and $3/2$, where ${\Omega}^{+}$, ${\Lambda}^{+}$ and ${\Sigma}^{+}$ are the quantum numbers associated with the projection of the total angular momentum ($\vec{J}^{+}$), the electron orbital ($\vec{L}^{+}$) and the electron spin ($\vec{S}^{+}$) angular momenta on the molecule axis, respectively. The $\textrm{X}^{+}\,{}^{2}\Pi_{3/2}$ and $\textrm{X}^{+}\,{}^{2}\Pi_{1/2}$ states are spectrally separated by the spin--orbit splitting $\Delta\tilde{\nu}_{\textrm{so}}$, and $\textrm{X}^{+}\,{}^{2}\Pi_{3/2}$ is the ground state of the cation.

At the low rotational temperature of the jet-cooled sample ($\approx 15$~K, see below), centrifugal-distortion effects are small and the rotational level energies of \ce{HC4I} are described in good approximation by the standard expression \cite{Bauder2011}
\begin{equation}
E''_{\textrm{rot}}(\textrm{X}\,{}^{1}\Sigma^{+})/(hc) = \tilde{B}_{0} J'' (J''+1),
\label{eq:rotneutral}
\end{equation}
where $J'' = 0,1,2, \dots$ is the total angular-momentum quantum number. The rotational constant $\tilde{B}_{0}$ is estimated to be about 0.0257~cm$^{-1}$ from the bond lengths in diacetylene, \ce{HC4H}, \cite{Matsumura2006, Matsumura2007} and iodoacetylene, \ce{HC2I}\ \cite{Andresen1989}. 

The rotational level energies of \ce{HC4I+} in the $\textrm{X}^{+}\,{}^{2}\Pi_{{\Omega}^{+}}$ vibrationless ground state are given by \cite{Woerner2011}
\begin{multline}
E^{+}_{\textrm{rot}}(\textrm{X}\,{}^{2}\Pi_{{\Omega}^{+}})/(hc) \\
= \tilde{B}^{+}_{0} \left[\left(J^{+}-\frac{1}{2}\right)\left(J^{+}+\frac{3}{2}\right)
\pm\sqrt{\left(J^{+}-\frac{1}{2}\right)\left(J^{+}+\frac{3}{2}\right)+\left(\frac{\tilde{A}}{2\tilde{B}^{+}_{0}}-1\right)^2}\right],
\label{eq:rotcation}
\end{multline}
where $\tilde{A}$ is the spin--orbit-coupling constant and the positive (negative) sign in front of the square root refers to the ${\Omega}^{+} = 1/2$ (3/2) spin--orbit component. The rotational constant $\tilde{B}^{+}_{0}$ is assumed to be the same for both spin--orbit states.  $J^{+} = {\Omega}^{+}, {\Omega}^{+}+1, {\Omega}^{+}+2, \dots$ designates the total angular momentum quantum number. The rotational structure of the photoelectron spectrum is calculated using
\begin{equation}
hc\tilde{\nu} = E_{\textrm{I}} + E_{\textrm{rot}}^{+} - E_{\textrm{rot}}'',
\label{eq:midp}
\end{equation}
where $E_{\textrm{I}}$ represents the midpoint between the ${}^{2}\Pi_{3/2}$ and ${}^{2}\Pi_{1/2}$ spin--orbit components.

The rotational line intensities within each band are obtained using the orbital ionisation model developed by Buckingham, Orr and Sichel \cite{Buckingham1970}. In this model, the experimentally observed spectral intensities are related to the molecular orbital out of which ionisation occurs. In the molecule-fixed frame, the angular part of this orbital is given by a linear combination of spherical harmonics $\sum_{l'' \geq \left|{\lambda}''\right|} c_{l''} Y_{l'' {\lambda}''}$ centred at the centre of mass, with $l''$ and ${\lambda}''$ being the quantum numbers associated with the orbital angular momentum and its projection onto the internuclear axis, respectively. The rotational line intensities $\sigma_{\textrm{tot}}(J'',J^{+})$ of the photoelectron spectrum are calculated using \cite{Buckingham1970, Merkt1993}
\begin{equation}
\sigma_{\textrm{tot}}(J'',J^{+}) = f'' G  \sum_{l'' \geq {\lambda}''} \frac{1}{2 l'' + 1} \left(2J^{+}+1\right) \left(2S''+1\right) Q_{l''} a_{l''},
\label{eq:intensities}
\end{equation}
where $G$ is a constant proportional to the Franck-Condon factors and ${\lambda}''$ is the number of orbital nodal planes containing the internuclear axis, i.e., ${\lambda}'' = 1$ for ionisation out of a $\pi$ orbital. 
A Boltzmann factor, $f'' = (2J''+1)\exp\left(-E''_{\textrm{rot}}/\left(k_\textrm{B} T_{\textrm{rot}}\right)\right)$ with $k_\textrm{B}$ and $T_{\textrm{rot}}$ being the Boltzmann constant and the rotational temperature, respectively, accounts for the population of rotational energy levels in the supersonic beam. 
The factor $a_{l''}$ in Eq.~(\ref{eq:intensities}) stands for a sum of radial matrix elements and the factor $Q_{l''}$ is the corresponding angular integral \cite{Buckingham1970},
\begin{equation}
Q_{l''} = 
\begin{pmatrix}
    S^{+} & \frac{1}{2} & S'' \\
    {\Sigma}^{+} & -{\Delta\Sigma} & -{\Sigma}''
\end{pmatrix}^2
\sum_{{\chi}} \left(2{\chi}+1\right)
\begin{pmatrix}
    l'' & \frac{1}{2} & {\chi} \\
    -{\lambda}'' & {\Delta\Sigma} & -{\Delta\Omega}
\end{pmatrix}^2
\begin{pmatrix}
    J^{+} & {\chi} & J'' \\
    -{\Omega}^{+} & {\Delta\Omega} & {\Omega}''
\end{pmatrix}^2,
\label{eq:Q}
\end{equation}
with ${\chi} = l''\pm\frac{1}{2}$, ${\Delta\Sigma} = {\Sigma}^{+}-{\Sigma}''$ and ${\Delta\Omega} = {\Omega}^{+}-{\Omega}''$. The Wigner 3-$j$ symbols in Eq.~(\ref{eq:Q}) imply the photoionisation selection rule ${\Delta J} = J^{+}-J'' = l''+1/2, l''-1/2, \dots, -l''-1/2$. In the analysis, the coefficients $a_{l''}$ are used as adjustable parameters.

\begin{figure}[ht!]
\centering
\includegraphics[width=0.57\textwidth]{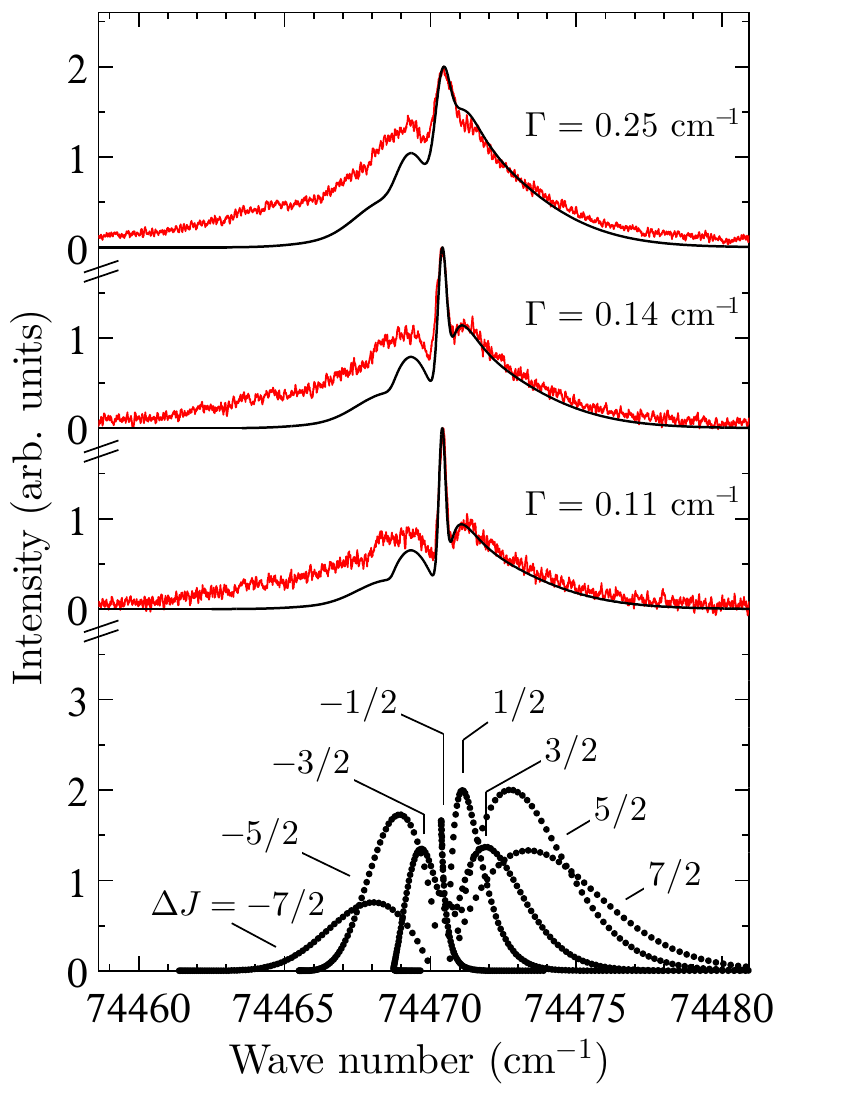}
\caption{\label{fig:32} Experimental PFI-ZEKE photoelectron spectra (red traces) of the $\textrm{X}^{+}\,{}^{2}\Pi_{3/2} \leftarrow \textrm{X}\,{}^{1}\Sigma^{+}$ $0^{0}_{0}$ transition of \ce{HC4I} (band maximum at $\tilde{\nu}=74470.7$~cm$^{-1}$) recorded using different steps of the PFI sequence and calculated spectra (black traces) obtained using Eqs.~(\ref{eq:rotneutral})--(\ref{eq:Q}). The calculated spectra are obtained by a convolution of the calculated stick spectrum (dots, only relative intensities of rotational lines are given) with a Gaussian line-shape function of full width at half maximum $\Gamma$. The traces are vertically offset for clarity.}
\end{figure}

\begin{figure}[ht!]
\centering
\includegraphics[width=0.57\textwidth]{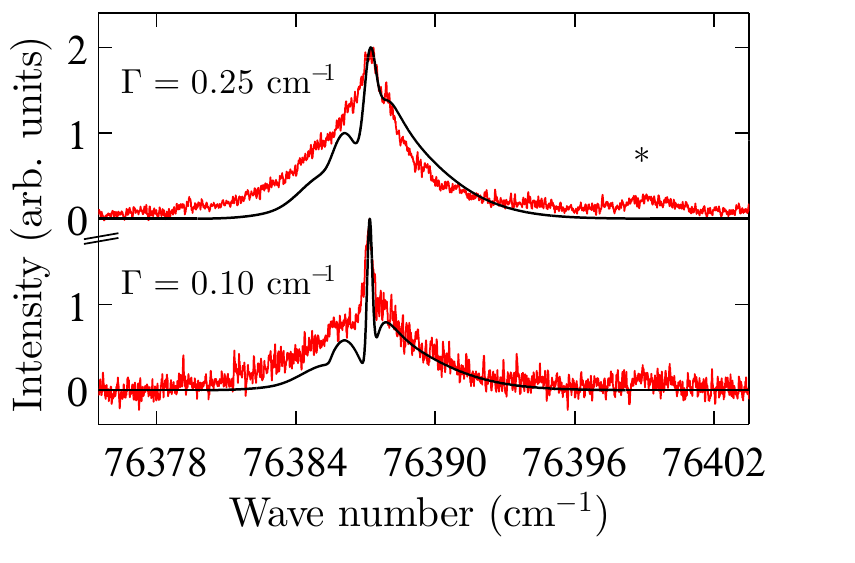}
\caption{\label{fig:12} Experimental PFI-ZEKE photoelectron spectra (red traces) of the $\textrm{X}^{+}\,{}^{2}\Pi_{1/2} \leftarrow \textrm{X}\,{}^{1}\Sigma^{+}$ $0^{0}_{0}$ transition of \ce{HC4I} (band maximum at $\tilde{\nu}=76387.4$~cm$^{-1}$) recorded using different steps of the PFI sequence and calculated spectra (black traces) determined using Eqs.~(\ref{eq:rotneutral})--(\ref{eq:Q}); labelling as in Fig.~\ref{fig:32}. The weak band marked with an asterisk (band maximum at $\tilde{\nu}=76399$~cm$^{-1}$) is assigned to the $\textrm{X}^{+}\,{}^{2}\Pi_{3/2} \leftarrow \textrm{X}\,{}^{1}\Sigma^{+}$ $3^{1}_{0}$ transition.}
\end{figure}

Figs.~\ref{fig:32} and \ref{fig:12} show experimental and calculated photoelectron spectra of the $0^{0}_{0}$ transitions from the $\textrm{X}\,{}^{1}\Sigma^{+}$ rovibronic ground state of \ce{HC4I} to the lower and upper spin--orbit components of the $\textrm{X}^{+}\,{}^{2}\Pi$ state of \ce{HC4I+}, respectively. The electric-field-induced shifts of the ionisation thresholds and the corresponding full widths at half maximum $\Gamma$ of the lines recorded with each pulse were estimated from a numerical simulation of the PFI process \cite{Hollenstein2001}. Depending on the PFI sequence and the pulse selected, a spectral resolution between $\Gamma = 0.10$~cm$^{-1}$ and 0.25~cm$^{-1}$ was achieved. The spectral line positions and intensities of the photoelectron spectra were calculated using Eqs.~(\ref{eq:rotneutral})--(\ref{eq:Q}). The resulting stick spectra were convoluted with Gaussian line-shape functions of full width at half maximum $\Gamma$, and compared to the experimental spectra. The parameters of the model, $A$, $E_{\textrm{I}}$, $T_{\textrm{rot}}$, $\tilde{B}^{+}_{0}$ and $a_{l''}$ were successively refined until optimal agreement between observed and calculated spectra was obtained. 

The photoelectron spectra consist of a narrow, intense central line with full width at half maximum limited by the experimental resolution and broader rotational side branches. The central line could unambiguously be assigned to the $\Delta J = -1/2$ branch, indicating that $\tilde{B}_{0}^{+} > \tilde{B}_{0}$. 
Its sharpness also enabled the precise determination of the first adiabatic ionisation energies $E^{\textrm{ad}}_{\textrm{I}}$ of \ce{HC4I}, corresponding to the transition from the $J'' = 0$ ground state of the neutral to the $J^{+} = 3/2$ and the $J^{+} = 1/2$ rotational level of the X${}^{+}\,{}^{2}\Pi_{3/2}$ and X${}^{+}\,{}^{2}\Pi_{1/2}$ spin--orbit states, respectively.

\begin{table}[hbt]
\centering
\caption{Adiabatic ionisation energy, $E^{\textrm{ad}}_{\textrm{I}}$, corresponding to ionisation to the $\Omega^{+}=3/2$ spin--orbit component, spin--orbit coupling, $\Delta\tilde{\nu}_{\textrm{so}}$, of the $\textrm{X}^{+}\,{}^{2}\Pi$ vibrationless ground state of \ce{HC4I+}, positions, $\tilde{\nu}_{\textrm{vib,}i}$, of vibrationally excited states of the X$^{+}\,{}^{2}\Pi$ state of \ce{HC4I+}, and wave numbers relative to the vibrationless X$^{+}\,{}^{2}\Pi_{3/2}$ and X$^{+}\,{}^{2}\Pi_{1/2}$ states. Values determined in this work are compared to results from earlier He I photoelectron spectroscopic studies \cite{Heilbronner1974, Maier1980}. All values are given in cm$^{-1}$.}\label{tab:compresults}
\begin{threeparttable}
\begin{tabular}{ccccccc}
\toprule[1.0pt]
 & this work & $\Delta$(${}^{2}\Pi_{3/2}$) & $\Delta$(${}^{2}\Pi_{1/2}$) & Ref. \cite{Heilbronner1974} & Ref. \cite{Maier1980} & mode\\
\midrule[1.0pt]
$E^{\textrm{ad}}_{\textrm{I}}/(hc)$ & 74470.7(2) & & & 74525(161) & 74848(81) & \\
$\Delta\tilde{\nu}_{\textrm{so}}$         & 1916.7(4) & & & 2017      & 2017     & \\
\midrule[1.0pt]
$\tilde{\nu}_{\textrm{vib},1}$~\tnote{a} & 76399.0(10) & 1928.4 & 11.7 &      & 2000 & $3^{1}_{0}$\\
$\tilde{\nu}_{\textrm{vib},2}$~\tnote{a} & 76573.6(20) & 2103.0 & 186.3 & 2100 &      & $2^{1}_{0}$\\
$\tilde{\nu}_{\textrm{vib},3}$~\tnote{a} & 76954.3(10) & 2483.7 & 567.0 &      &      & \\
\bottomrule[1.0pt]
\end{tabular}
\begin{tablenotes}
\item[a] For comparison, the fundamental vibrational wave numbers of the X$\,{}^{1}\Sigma^{+}$ electronic ground state of \ce{HC4I} are \cite{Christensen1969, Minasso1971}: $\tilde{\nu}_{1}$(C--H stretch) = 3332~cm$^{-1}$, $\tilde{\nu}_{2}$(upper (C$\equiv$C)$_2$ stretch) = 2211~cm$^{-1}$, $\tilde{\nu}_{3}$(lower (C$\equiv$C)$_2$ stretch) = 2060~cm$^{-1}$, $\tilde{\nu}_{4}$(C--C stretch) = 1025~cm$^{-1}$, $\tilde{\nu}_{5}$(C--I stretch) = 362~cm$^{-1}$, $\tilde{\nu}_{6}$(C$\equiv$C--H bend) = 623~cm$^{-1}$, $\tilde{\nu}_{7}$((C$\equiv$C)$_2$ asymmetric bend) = 473~cm$^{-1}$, $\tilde{\nu}_{8}$((C$\equiv$C)$_2$ symmetric bend) = 357~cm$^{-1}$, $\tilde{\nu}_{9}$(C$\equiv$C--I bend) = 106~cm$^{-1}$. The spin--orbit splitting is given as a positive interval. The value of the spin--orbit coupling constant is negative.
\end{tablenotes}
\end{threeparttable}
\end{table}

The values obtained for the first adiabatic ionisation energy of \ce{HC4I} and the spin--orbit splitting between the ${}^{2}\Pi_{3/2}$ and ${}^{2}\Pi_{1/2}$ components of the X${}^{+}$ state of \ce{HC4I+} are compared with the results of earlier He I photoelectron spectroscopic studies \cite{Heilbronner1974, Maier1980} in Table~\ref{tab:compresults}. 

Because our experiment is only sensitive to the change of the rotational constant $\Delta \tilde{B} = \tilde{B}^{+}_{0}-\tilde{B}_{0}$, the estimated value for $\tilde{B}_{0} = 0.0257$~cm$^{-1}$ was kept constant in the optimisation procedure, yielding $\Delta \tilde{B} = 0.0008(4)$~cm$^{-1}$. A rotational temperature of 15~K was found to reproduce the experimentally observed intensity distributions best.  In the future, more accurate rotational constants of both the ion and the neutral could thus be derived by an independent determination of either $\tilde{B}^{+}_{0}$ or $\tilde{B}_{0}$. 

The intensities of the branches with $\Delta J \geq -1/2$ could be almost perfectly reproduced by the calculations using $\left|{\lambda}''\right| = 1$ and $a_{l''} = 0.1, 0.1,$ and 1 for $l'' = 1, 2$ and 3, respectively, which indicates a dominant f contribution to the highest occupied molecular orbital of \ce{HC4I}. The intensities of the $\Delta J < -1/2$ branches were systematically underestimated by the calculations. This observation can be attributed to the effects of rotational channel interactions, which tend to enhance the intensities of rotational branches with negative $\Delta J$ values compared to those with positive $\Delta J$ values in PFI-ZEKE photoelectron spectra (see discussion in Ref.~\cite{Merkt1993}).

\subsection{Vibrational Levels of the X$^{+}\,{}^{2}\Pi$ State of \ce{HC4I+}}

\begin{figure}
\begin{center}
\begin{minipage}{130mm}
\subfigure[]{
\resizebox*{6.2cm}{!}{\includegraphics{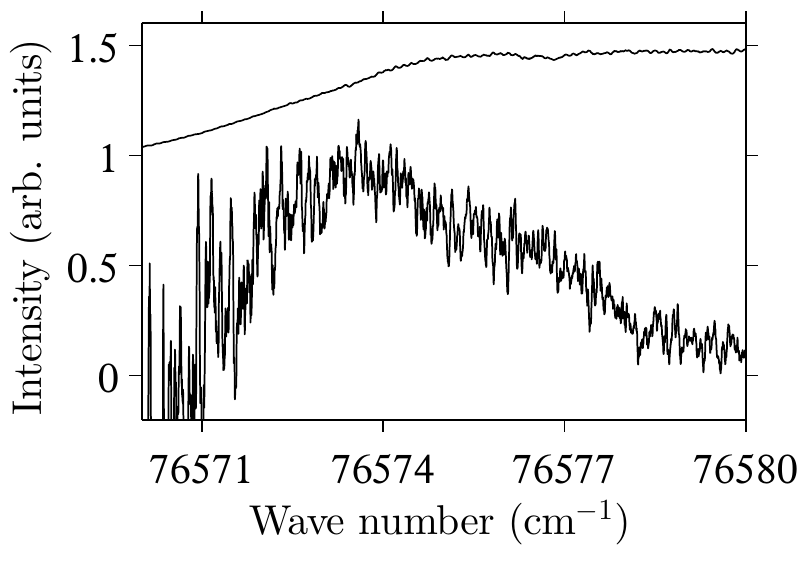}}}\label{fig:asymstr}%
\subfigure[]{
\resizebox*{6cm}{!}{\includegraphics{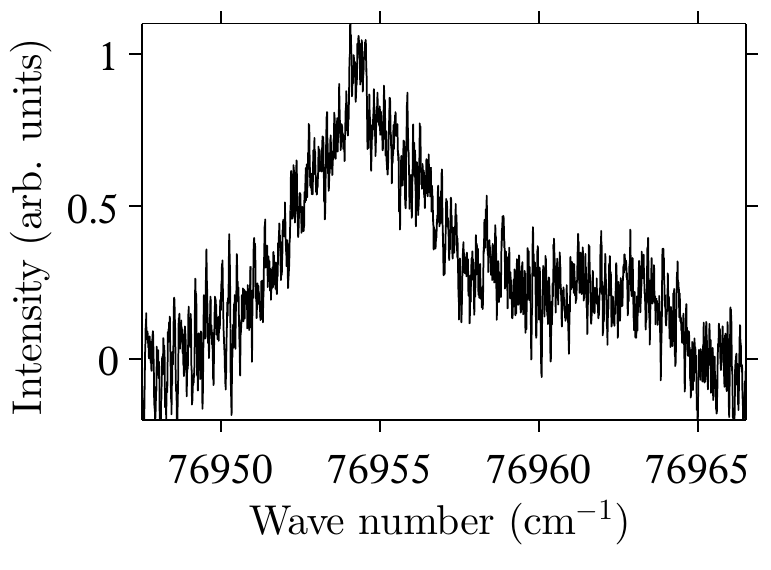}}}\label{fig:comb}%
\caption{PFI-ZEKE photoelectron spectra of \ce{HC4I+} above the origin of the $\textrm{X}^{+}\,{}^{2}\Pi_{1/2} \leftarrow \textrm{X}\,{}^{1}\Sigma^{+}$ transition. PFI-ZEKE photoelectron spectra recorded from different pulses were corrected for the field-induced shifts and then summed to give photoelectron spectra with a resolution of $\Gamma = 0.25$~cm$^{-1}$. (a) Spectrum of the $\textrm{X}^{+}\,{}^{2}\Pi_{3/2} \leftarrow \textrm{X}\,{}^{1}\Sigma^{+}$ $2^{1}_{0}$ transition (lower trace) with intensities normalised to the VUV laser intensity (upper trace, vertically offset by +1). (b) Spectrum of an unassigned transition (see text).}%
\label{fig:vibmodes}
\end{minipage}
\end{center}
\end{figure}

Transitions to three excited vibrational levels of the X$^{+}\,{}^{2}\Pi$ state of \ce{HC4I+} were also observed. The PFI-ZEKE photoelectron spectra corresponding to these transitions are presented in Figs.~\ref{fig:12} (band marked with an asterisk) and \ref{fig:vibmodes}. Compared to the $0^{0}_{0}$ transitions, these transitions are much weaker. For this reason, PFI sequences with larger field steps had to be used to record the spectra displayed in Fig.~\ref{fig:vibmodes}, which reduced the spectral resolution and made assignments based on the rotational branch structure of the photoelectron spectra impossible. The measurement of the spectrum shown in Fig.~\ref{fig:asymstr} was further complicated by a drop in VUV laser intensity below 76575~cm$^{-1}$.

\ce{HC4I+} has 13 vibrational modes, 5 modes of $\sigma^{+}$ symmetry and 4 doubly degenerate modes of $\pi$ symmetry. In the absence of rovibronic interactions, only the $\sigma^{+}$ vibrational levels of the cation can be excited from the vibronic ground state of the neutral molecule. Previous photoelectron spectroscopic studies of \ce{HC2I+} showed that transitions to vibrational levels of $\pi$ symmetry, such as the fundamental of the C$\equiv$C--I bending mode, can become weakly allowed through vibronic interactions \cite{Gans2013}. 

The transitions observed in the photoelectron spectrum of \ce{HC4I} could in principle be transitions to vibrational levels of either of the two spin--orbit states. In Table~\ref{tab:compresults}, the band centres are given relative to both the ${\Omega}^{+} = 3/2$ and $1/2$ band origins. The band with a maximum at 76399.0~cm$^{-1}$ must correspond to a vibrationally excited level of the X$^{+}\,{}^{2}\Pi_{3/2}$ state, because the energetic difference to the origin of the ${}^{2}\Pi_{1/2}$ state is too small for a vibrational excitation. We assign this band to a transition to the $3^{1}$ (lower (C$\equiv$C)$_2$ stretch) level. 

The band at 76573.6~cm$^{-1}$ corresponds to a vibrational wave number of 186.3~cm$^{-1}$ for the ${\Omega}^{+} = 1/2$ component, which rules out all vibrational levels except the fundamental of the C$\equiv$C--I bending mode ($\pi$ vibrational symmetry).
The vibrational wave number for the ${\Omega}^{+} = 3/2$ component is 2103.0~cm$^{-1}$ and lies in the range expected for the upper of the two (C$\equiv$C)$_2$ stretching modes ($\sigma^+$ vibrational symmetry). We favour the latter assignment because it corresponds to an allowed transition that is expected to have a nonzero Franck-Condon factor because ionisation weakens the triple bonds (see Section~\ref{discussion}). 

These two assignments are in agreement with those obtained in earlier photoelectron spectroscopic studies, but the fundamental vibrational wave numbers derived here are more precise (see Table~\ref{tab:compresults}). The decrease in these two vibrational wave numbers with respect to those of \ce{HC4I} is consistent with the weakening of the C$\equiv$C bonds upon ionisation (see Sec.~\ref{discussion}).

The assignment of the band at 76954.3~cm$^{-1}$ is not straightforward. The band corresponds to a wave number of 567~cm$^{-1}$ for the ${\Omega}^{+} = 1/2$ component, which corresponds to the wave-number range of the fundamental vibration of the modes $\nu_7$ and $\nu_8$ of neutral \ce{HC4I} (see Table~\ref{tab:compresults}) and \ce{HC4I+}. However, these modes have $\pi$ symmetry. The vibrational wave number for the ${\Omega}^{+} = 3/2$ component is 2483.7~cm$^{-1}$, which is too high for any of the fundamental wave numbers of \ce{HC4I+}. Combination bands or overtones cannot be ruled out, so that it is not possible to give an unambiguous assignment for this band.

\section{Discussion}
\label{discussion}

A qualitative understanding of the photoelectron spectrum of \ce{HC4I} can be reached in the realm of the three-state model described by Eq.~(\ref{eq:Hamiltonian}), as already pointed out in Ref.~\cite{Heilbronner1974}. Rather than adapting the model parameters to get the best possible agreement with the level positions measured in this work and in previous studies, we
fixed all parameters to experimentally measured ionisation energies and to interactions derived from the spectra of other molecules. We used (i) a value of 84886~cm$^{-1}$ for the ionisation energy $E_{\textrm{I,X}}/(hc)$ of the iodine atom, which corresponds to the difference between the centres of gravity of the three spin--orbit components of the $^3$P ground state of I$^+$ and the two spin--orbit components of the $^2$P ground state of I\ \cite{Gruetter2012}, (ii) a value of 91953.5~cm$^{-1}$ for the ionisation energy $E_{\textrm{I,\ce{HC2H}}}/(hc)$ of acetylene \cite{Rupper2004}, (iii) a value of 7632~cm$^{-1}$ for the charge-transfer coupling parameter $V_{\textrm{I-CC}}$ derived from the photoelectron spectrum of HC$_2$I\ \cite{Gans2013}, and (iv) a value of 9840~cm$^{-1}$ for the charge-transfer coupling parameter $V_{\pi\pi}$ extracted from the photoelectron spectrum of diacetylene \cite{Smith1967,Brogli1973}, as detailed in Table~\ref{tab:modeloutput}. 

The eigenvalues $E_{\textrm{I},i=1-3}$ of the matrix~(\ref{eq:Hamiltonian}), correspond to the energetic positions of the three lowest ${}^{2}\Pi$ electronic states of HC$_4$I$^+$ with respect to the X$\,{}^{1}\Sigma^{+}$ electronic ground state of the neutral molecule, neglecting spin--orbit coupling. 
The elements of the eigenvector $\vec{v} = \left[c_{\rm{I}},\,c_{\rm{CC(1)}},\,c_{\rm{CC(2)}}\right]^{\textrm{T}}$ associated with each electronic eigenstate designate the halogen and triple-bond character of the electron hole generated by photoionisation. Neglecting the contribution from the \ce{HC4} chain, the spin--orbit splittings are obtained as $\Delta\tilde{\nu}_{\textrm{so}} = \left|c_{\textrm{I}}\right|^2 |\tilde{A}_{\textrm{I}}|$, where $c_{\textrm{I}}$ represents the eigenvector element associated with the iodine atom, and $\tilde{A}_{\textrm{I}} = -5069$~cm$^{-1}$ is its spin--orbit coupling constant.

\begin{table}[hbt]
\centering
\caption{Experimental (normal print) and calculated (italic print) values of the ionisation energies, $E_{\textrm{I}}/(hc)$, and spin--orbit splittings, $\Delta\tilde{\nu}_{\textrm{so}}$, in linear acetylenic systems. Experimental data are taken as the centres of gravity of the ${}^{2}\Pi_{3/2}$ and ${}^{2}\Pi_{1/2}$ ionisation thresholds obtained from He I and PFI-ZEKE photoelectron spectroscopic measurements, as indicated. Calculated values are obtained from two- or three-state charge-transfer models, see main text.}\label{tab:modeloutput}
\begin{threeparttable}
\begin{tabular}{lllllll}
\toprule[1.0pt]
 & \multicolumn{3}{l}{$E_{\textrm{I}}/(hc)$ (in cm$^{-1}$)} & \multicolumn{3}{l}{$\Delta\tilde{\nu}_{\textrm{so}}$ (in cm$^{-1}$)}   \\
Molecule & X$^{+}\,{}^{2}\Pi$ & A$^{+}\,{}^{2}\Pi$ & B$^{+}\,{}^{2}\Pi$ & X$^{+}\,{}^{2}\Pi$ & A$^{+}\,{}^{2}\Pi$ & B$^{+}\,{}^{2}\Pi$ \\
\midrule[1.0pt]

\ce{HC4H} & 82107\ \cite{Smith1967} & 101787\ \cite{Smith1967, Brogli1973} &  &  &  & \\
          & \textit{82114} & \textit{101793}             &  &  &  & \\

\ce{HC2I} & 79925\ \cite{Gans2013}  & 96746\ \cite{Allan1977} &  & 3257\ \cite{Gans2013}  & 2016\  \cite{Allan1977} & \\ 
          & \textit{80009} & \textit{96830} &  & \textit{3599} & \textit{1470}  & \\ 

\ce{HC4I} & 75429~\tnote{a} & 90616\ \cite{Heilbronner1974}\,\tnote{b}  & 101626\ \cite{Heilbronner1974},\tnote{b}  & 1917~\tnote{a} & 2984\ \cite{Heilbronner1974}\,\tnote{b}  & 169~\tnote{c} \\
           & \textit{77448} & \textit{87862} & \textit{103483} & \textit{2124} & \textit{2495} & \textit{450} \\
\bottomrule[1.0pt]
\end{tabular}
\begin{tablenotes}
\item[a] this work
\item[b] The ionisation energy corresponding to the formation of the ${\Omega}^{+}=1/2$ component of the B$^+$ state in Table I of Ref.~\cite{Heilbronner1974} (14.9 eV) must be that corresponding to the formation of the C$^+$ state. For the comparison, we used the experimental ionisation energies listed in Table I of Ref.~\cite{Heilbronner1974}, which differ from those listed in the last column of Table VII  of the same reference.
\item[c] This value results from the fact that, within the three-state model, the sum of the spin--orbit splittings must be equal to the atomic spin--orbit coupling constant $\tilde{A}_{\rm I}$.
\end{tablenotes}
\end{threeparttable}
\end{table}

Considering the simplicity of the model and the fact that no parameters have been adjusted, the comparison with experimental data (Table~\ref{tab:modeloutput}) can be considered satisfactory on a qualitative level. Calculated and experimental values of the ionisation energies all agree within 0.35 eV. Calculated and measured spin--orbit splittings all agree within 500~cm$^{-1}$ and follow the same trend, i.e., a maximal value for the A$^+$ state and a small value for the B$^+$ state. The spin--orbit splittings of the X$^+$ and A$^+$ states are large and imply an almost complete quenching of the Renner--Teller effect, as in HC$_2$I$^+$ \cite{Gans2013}. The observed spin--orbit splitting for the X$^{+}\,{}^{2}\Pi$ state indicates that the electron hole is delocalised. For the X$^+$ state, the calculations give values of 0.419, 0.398 and 0.183 for the squared coefficients of the wavefunction corresponding to the p$_{\pi}$ orbital of the iodine atom ($|c_{\rm{I}}|^2$), the acetylenic unit adjacent to the I atom ($|c_{\rm{CC(1)}}|^2$), and the acetylenic unit adjacent to the H atom ($|c_{\rm{CC(2)}}|^2$), respectively. This delocalisation, in turn, implies a strengthening of the C--I bond and a weakening of the triple bonds compared to the neutral, which are also reflected in the values of the vibrational wave numbers and rotational constants obtained from the high-resolution photoelectron spectra presented in Section~\ref{results}. 

\section{Conclusions}

New information on the electronic ground state of the \ce{HC4I+} radical cation has been obtained using photoelectron spectroscopy. The high resolution that was achieved by PFI-ZEKE photoelectron spectroscopy (about 0.1~cm$^{-1}$) enabled the partial resolution of the rotational structure despite the small values (less than 0.03 cm$^{-1}$) of the rotational constants of the neutral and cationic states. 
This advantage was exploited to determine the first adiabatic ionisation energy of HC$_4$I, the spin--orbit splitting and vibrational intervals in the X$^+$ $^2\Pi$ ground electronic state of HC$_4$I$^+$ with improved accuracy over previous studies. 

Given the simplicity of the model (originally from Ref.\cite{Heilbronner1974}, but with a different parametrisation) used in the discussion of the experimental data and the fact that no parameters were adjusted, it is clear that the model predictions are less accurate than predictions by state-of-the-art {\it ab-initio} quantum-chemical methods would be in systems with a similar number of electrons, i.e. 77 electrons in \ce{HC4I+}. However, the model offers the advantages of being simple and of providing qualitatively correct trends for the ionisation energies, spin--orbit intervals, rotational constants and vibrational intervals. The model also nicely links the charge redistribution taking place upon ionisation to the ionisation energies of the subunits along the molecular chain. 

\section*{Acknowledgements}

The authors thank Urs Hollenstein (ETH Zurich) for experimental help and guidance in the operation of the VUV laser system. This work was supported financially by the Swiss National Science Foundation under Project No. 200020-159848.

\appendices

\section{Chemical Synthesis of HC$_{4}$I}

\subsection{General information}
\begin{figure}[ht!]
\centering
\includegraphics[width=0.9\textwidth]{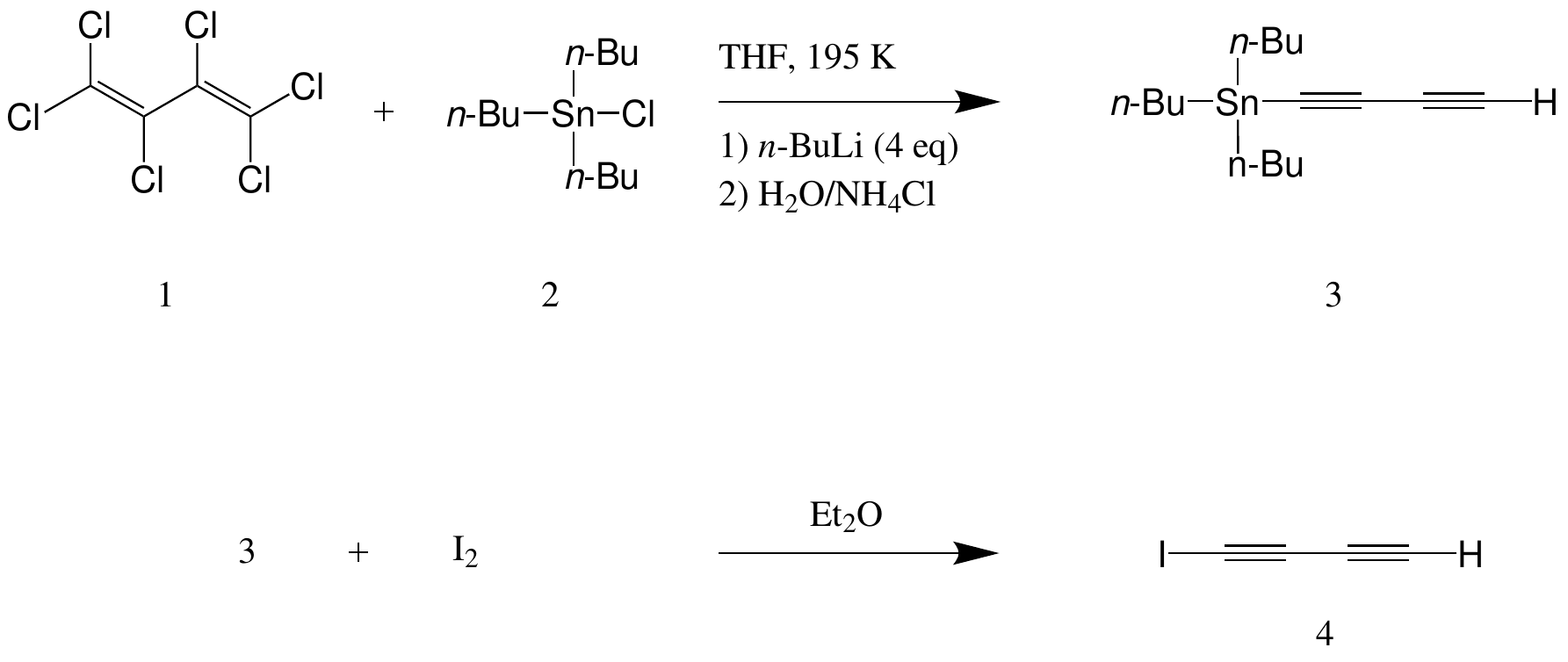}
\caption{\label{fig:synthesis} Reaction scheme of the two-step synthesis of iododiacetylene \textbf{4} via buta-1,3-diyn-1-yltributylstannane~\textbf{3}.}
\end{figure}

The reaction schemes used to synthesize \ce{HC4I} are presented in Fig.~\ref{fig:synthesis}. Chemicals: hexachloro-1,3-butadiene (ABCR, 97~\%), \textit{n}-butyllithium (\textit{n}-BuLi, Sigma-Aldrich, 1.6~M solution in \textit{n}-hexane), tributylchlorostannane (ABCR, 96~\%), tetrahydrofuran (THF, Acros Organics, 99.5~\% extra dry over molecular sieve), iodine (Aldrich, 99.8~\%), diethyl ether (\ce{Et2O}, Acros Organics, 99.5~\% extra dry over molecular sieve), nitrogen gas (dried by passing through phosphorus pentoxide). Chemical analysis: electron-ionisation GC-MS, $^1$H-NMR.

\subsection{Synthesis of Buta-1,3-diyn-1-yltributylstannane~\textbf{3}}
THF (250~mL) was placed inside a 1-L three-necked round-bottomed flask held at 195~K. While stirring, a \textit{n}-BuLi solution (250~mL, 0.4~mol) was added under nitrogen atmosphere over a period of 1~hour. After stirring for 30~minutes at 195~K, hexachloro-1,3-butadiene \textbf{1} (30~g, 0.12~mol) was added during 1.5~hours. The reaction vessel was wrapped in aluminium foil, the cooling bath was removed and the solution (brown colour) was stirred for 12~hours. The reaction mixture was then cooled to 195~K again, and tributylchlorostannane \textbf{2} (32.6~g, 0.1 mol) was added over a period of 2~hours. After stirring for another 30~minutes at 195~K, the cooling bath was removed and the mixture was allowed to warm up to room temperature while stirring for 12~hours. After the addition of a saturated aqueous ammonium chloride solution (150 mL) at room temperature, the mixture was extracted with \ce{Et2O} to yield 24.0~g of a brown oil with a content of 52~\% buta-1,3-diyn-1-yltributylstannane~\textbf{3}, 45~\% hexachlorobutadiene~\textbf{1}, and 3~\% tetrabutylstannane as a by-product. The product \textbf{3} was diluted in \ce{Et2O} and characterised by electron-ionisation GC-MS. The molecular ion peak M$^{+}$ was not detected but the fragment [M~$-$~\ce{C4H9}]$^{+}$ was observed. The following intensities were measured relative to the main peak $m/z = 168.93$ (100~\%): 279.03 (46.0~\%), 280.08 (29.6~\%), 281.06 (41.1~\%), 282.09 (25.5~\%), 283.06 (89.2~\%), 284.09 (10.94~\%), 285.06 (15.4~\%), 287.05 (17.0~\%). $^1$H-NMR (300~MHz, \ce{CDCl3}): $\delta$, ppm 1.96 (1H, s), 1.53--1.62 (12H, m), 1.33--1.37 (6H, m), 0.92 (9H, t, $J=7.2$~Hz). Here, only chemical shifts ascribed to \textbf{3} are given.

\subsection{Synthesis of Iododiacetylene~\textbf{4}}
Buta-1,3-diyn-1-yltributylstannane \textbf{3} (13~g, 20~mmol) was dissolved in \ce{Et2O} (10~mL) in a 250~mL two-necked round-bottomed flask connected to a cold trap. Under nitrogen atmosphere at 195~K, iodine (6.4~g) dissolved in \ce{Et2O} (60~mL) was added through a septum stopper over a period of 5 minutes using a syringe. The brownish solution was stirred for 12~hours at 195~K. 
The volatile parts of the reaction mixture, including the iododiacetylene product \textbf{4}, were transferred into a cold trap held at 218~K (using an immersion cooler) by trap-to-trap condensation under vacuum. After 5~hours, the valve between the flask and the cold trap was closed and the \ce{Et2O} in the trap was removed by continuous pumping of the mixture over 3~days while keeping the trap cooled to 218~K. The product \textbf{4} was obtained as a white to slightly yellow solid at 218~K. It was stored inside a long glass capillary at 77~K to prevent polymerisation at higher temperatures \cite{KlosterJensen1966}. Because of the instability of \textbf{4}, no NMR measurements were performed. The purity of \textbf{4} was estimated as 90~\% by electron-ionisation GC-MS; a 10~\% impurity was identified as \ce{Et2O}. GC-MS: $m/z$ and corresponding relative intensities, 175.9 (100~\%), 126.9 (21.0~\%), 48.99 (27.5~\%).\clearpage


\clearpage

\end{document}